# A Meta-evaluation of Scientific Research Proposals:

# Different Ways of Comparing Rejected to Awarded Applications

Lutz Bornmann (corresponding author),[i] Loet Leydesdorff,[ii] & Peter Van den Besselaar[iii]

**Abstract**

Combining different data sets with information on grant and fellowship applications submitted to two renowned funding agencies, we are able to compare their funding decisions (award and rejection) with scientometric performance indicators across two fields of science (life sciences and social sciences). The data sets involve 671 applications in social sciences and 668 applications in life sciences. In both fields, awarded applicants perform on average better than *all* rejected applicants. If only *the most preeminent rejected* applicants are considered in both fields, they score better than the awardees on citation impact. With regard to productivity we find differences between the fields: While the awardees in life sciences outperform on average the most preeminent rejected applicants, the situation is reversed in social sciences.

**Keywords**: grant allocation, peer review, bibliometric quality indicators, convergence validity and predictive validity, error, citation rate, h-index.

[i] ETH Zurich, Professorship for Social Psychology and Research on Higher Education, Zaehringerstr. 24, CH- 8092 Zurich, Tel.: +41 44 632 4825, Fax: +41 44 632 1283, E-mail: bornmann@gess.ethz.ch
[ii] Amsterdam School of Communications Research (ASCoR), University of Amsterdam, Kloveniersburgwal 48, 1012 CX Amsterdam; loet@leydesdorff.net; www.leydesdorff.net.
[iii] Science System Assessment Department, Rathenau Instituut, Anna van Saxenlaan 51, 2593 HW Den Haag, Email: p.vandenbesselaar@rathenau.nl & Department of Organization Sciences, Vrije Universiteit Amsterdam.
Authors in alphabetical order.

**Introduction**

The evaluation of research using scientometric indicators at the institutional and individual level emerged in the first half of the 1980s (Martin & Irvine, 1983; Moed *et al*., 1985). Before this date, Garfield (1972) had advocated the evaluation of journals using these quantitative measures and Narin (1976) had focused on the macro-evaluation of nations and disciplines. In the discussions about the development of science indicators (e.g., Elkana *et al*., 1978), however, the focus had remained on science at the macro-level. Merton (1979), for example, in his Foreword to Garfield's (1979) *Citation Indexing* did not even mention a potential conflict between quantitative and qualitative evaluations of research although he can be considered as the leading sociologist for studying the processes of peer review and recognition in science in the 1970s (e.g., Merton, 1968; Zuckerman and Merton, 1971).

The scientometric turn towards the institutional and even individual level in the first half of the 1980s generated immediately a source of potential conflicts with the scientific establishment which relies on peer review for quality control. Can indicators measure quality of researchers and research groups independently? Eventually, the scientometric community had to retreat to claiming the measurement of "impact" because "quality control" was considered the domain of peer review to be carried by the scientific community itself (Martin & Irvine, 1985). Peer review, originally used mainly by journals, had become increasingly a tool for project and programme evaluation by the then emerging research councils (Mulkay, 1976; De Haan, 1994). Whereas peer review is used by journals for both the selection and the improvement of submissions, one can expect its function to be very different in allocation decisions (Frey & Osterloh, 2002).

The results of an objectified method for quality control in terms of publications and citations cannot be expected to match one-to-one with those of content-based peer review although both are connected. One would expect a moderate correlation between bibliometric measures and content-based peer review at the aggregated level (Martin & Irvine, 1983; e.g., Nederhof & Van Raan, 1987). In fact, these associations between scientometric and peer-review-based evaluations were reported in the literature (see an overview in Daniel, Mittag & Bornmann, 2007). On the basis of a recent evaluation of selection processes at the Dutch Economics and Social Research Council (MaGW-NWO, in short: MaGW), however, Van den Besselaar and Leydesdorff (2007, 2009) noted that most of these evaluations focus on the net results of the selection process in terms of awarded and rejected applications, and do not deconstruct the internal mechanism of the selection. While these authors also found a statistically significant association between the scientometric evaluation and the outcome of the qualitative selections, this association became statistically significantly negative when comparing the awardees with the most preeminent rejected applicants in terms of publications and citations.

Van den Besselaar and Leydesdorff (2009) concluded that the "council's procedure operates well for identifying and discarding the tail of the distribution. However, within the top half of the distribution, neither the review outcomes nor past performance correlate positively with the decisions of the council." Furthermore, they typified the peer review process as the external organization of the funding agency: "The council has a large autonomy in prioritizing the applications. In this process both external reviews and performance indicators play only an auxiliary role" (Van den Besselaar and Leydesdorff, 2009). Similar results concerning the high performance of the most preeminent rejected applicants are reported by Melin and Danell (2006) when they examined the "publication histories" of the top eight percent of all applicants to the Swedish Foundation for Strategic Research (Stockholm) and found only slight mean differences in scientific productivity between awarded and rejected applicants. Also Hornbostel, Böhmer, Klingsporn, Neufeld, and von Ins (2009) report comparable findings for the Emmy Noether Programme of the German Research Foundation (DFG, Deutsche Forschungsgemeinschaft, Bonn) although they did not analyze certain subgroups of applicants but *all* applicants in the fields of medicine and physics with funding decisions between the years 2000 and 2006.

Bornmann, Wallon, and Ledin (2008a) analyzed the Long-Term Fellowship (LTF) and the Young Investigator (YI) programmes of the European Molecular Biology Organization (EMBO, Heidelberg, Germany). Both programmes aim to identify and support the best post doctoral fellows and young group leaders in the life sciences. They used the performance prior and subsequent to application as scientometric criteria for distinguishing between the entire groups of awarded and rejected applicants. If quantity and impact of research publications are used as a criterion for scientific achievement, the results of negative binomial regression models show that both EMBO programmes fund scientists who perform on a higher level than the rejected ones prior and subsequent to application (Bornmann, Wallon, & Ledin, 2008a).

The comparing of the awardees with the most preeminent rejected applicants by Van den Besselaar and Leydesdorff (2007, 2009) and the results of Melin and Danell (2006) point out that the approval and rejection decisions by the selection committees could have been different: If past performance of applicants were used as validity criterion, a considerable amount of rejected applicants could have been awarded and vice versa. Bornmann, Wallon, and Ledin (2008a) expanded on such approaches, such as those of Melin and Danell (2006) ("the top eight percent") and Van den Besselaar and Leydesdorff (2007, 2009) ("discarding the tail of distributions"), in order to compare selected groups of accepted and rejected applicants with each other. They determined the extent of "errors" due to over-estimation (type I errors among approved applicants) and under-estimation (type II errors among rejected applicants) of future scientific performance. Their statistical analyses point out that between 26% and 48% of the decisions made to award or reject an application show one of both error types. Whereas the validity of the selection process is given on average, a certain amount of decisions can be called "erroneous" in terms of citation impact.

This study emerged from the wish to compare and discuss the results of two studies concerning the relation between bibliometric performance indicators and selection decisions in different disciplinary contexts. Whereas the study of Bornmann, Wallon, and Ledin (2008a) is based on data from the area of the life sciences, Van den Besselaar and Leydesdorff (2007, 2009) focused on the social sciences. Using our previously analyzed data sets, we compare in this study the two data sets (MaGW and EMBO) by harmonizing the previously used methodologies and by thus making the results better comparable.

**Data sets and statistical procedures**

*Description of the data sets*

The MaGW data consist of 671 applications to the funding agency during the period 2003-2005 (see Table 1), covering the open competition and career programmes. Of these applications, 370 applications were in psychology and 301 in economics. Success rates were 32.2% (119) in psychology and 15.3% (46) in the case of economics. In Van den Besselaar and Leydesdorff (2007, 2009) a larger set of 1273 applications were used, covering all social science disciplines, including anthropology, communication studies, economics, law, political science and psychology. Of these 1273 applications, 275 (21.6%) were awarded. In this study we restrict the analysis to economics and psychology only, as the use of scientometric indicators for measuring research performance is here more accepted and even institutionalized than in other social science disciplines. It should be noted that applying the analysis reported in this paper to the whole set of 1273 applications yields similar results as in the cases of economics and psychology only.

[Table 1 to be placed here.]

The EMBO data sets involve 668 applications to the LTF programme in 1998 and 297 applications to the YI programme in the period 2001-2002 (see Table 1) (see a detailed description in Bornmann, Wallon, and Ledin, 2008a). Established in 1966, the LTF programme has gained an excellent reputation in the scientific community. The fellowships are awarded for a period of up to two years and are intended for advanced post doctoral research. The YI programme has been supporting outstanding young group leaders in the life sciences in Europe since 2000. The programme targets researchers who have established their first independent laboratories normally four years before the assessment in a European Molecular Biology Conference (EMBC) member state. Hundred-thirty (19.5%) of the LTF applications were awarded and 39 (13.1%) of the YI applications. All applications under these programmes can be considered as belonging to the field of molecular biology.

In summary, we use in this study (two times two =) four data sets from very different disciplinary areas. The areas are characterized by a different importance of scientometric indicators: in molecular biology the importance is valued higher than in psychology and economics. However, although both agencies operate in different disciplinary contexts, their selection processes are characterized by similar high rejection rates – an indication of their high renown.

In the MaGW case, three years of scientometric data were collected before and after the submission dates. Time windows were set to three years because this is the time window for the references provided by the applicants. The ex ante data was downloaded on February 7, 2007; the ex post data on June 1, 2009, using the Web of Science provided by Thomson Reuters (Philadelphia, PA). To make the EMBO data comparable to the MaGW data, the same publication windows before and after submission dates are considered in the analysis. That means in case of the LTF programme publications from 1996, 1997 and 1998 (ex ante) and from three years after application (ex post) are included in the analyses (Web of Science data, too). For the applicants to the YI programme we used in the analyses publications of two years prior to application (and those published in the application year, ex ante) and publications of three years subsequent to application (ex post). For the publications of the LTF applicants we have citations from publication year to the beginning of 2006; for those of the YI applicants from publication year to the beginning of 2007 (see here Bornmann, Wallon, and Ledin, 2008a).

The bibliometric data for both funding agencies were used to calculate numbers of publications (mean publication rates), numbers of citations (total citation counts and mean citation rates) and $h$ index values (Bornmann and Daniel, 2009) for every application. The data were also used for the calculation of multiple regression models.

Comparisons in this study are done both for the groups of awarded *versus* rejected applications and for the groups of awarded *versus* the best-rejected applications. The group of the best-rejected applications is defined as those rejected applications (the same number of applications as was awarded) with the highest mean citation rates for papers published prior to application. Thus, in the case of 119 awarded applications in psychology we used the subgroup of 119 rejected applications with the highest mean citation rates during the three years before application.[1] Table 1 summarizes the MaGW and EMBO data (number of applications and number of publications used in this study).

*Description of the statistical procedures*

In each data set, the difference between awarded and (best-)rejected applications are tested for statistical significance using the (non-parametric) Mann-Whitney test for two independent samples in the statistical package SPSS. Both groups are compared in two ways: by using scientometric indicators for the time prior (ex ante) and subsequent (ex post) to the application. With the indicators referring to the time prior to the application, the convergent validity of the agencies' selection decisions is tested. If there is a considerable association between indicators and decisions (both measure scientific quality in a quantitative and qualitative way, respectively), the selection decisions can be considered as *convergent valid*. With the indicators referring to the time subsequent to application the decisions' *predictive validity* is tested. In the latter case, we find an answer to the question, whether the agencies are able to select applications with the "best" future scientific performance.

Besides mean differences between awarded and (best-)rejected applicants the amount of "erroneous" decisions in both selection processes are calculated by using *h* index values. Type I and type II "errors" were defined by Bornmann, Wallon, and Ledin (2008a) as follows (see also Bornmann and Daniel, 2007a; Straub, 2008a, 2008b): a decision to award an application is considered a type I error if the *h* index of the applicant is lower than or equal to the median of the group of rejected applicants (over-estimation of an applicant's performance). Analogously, type II errors are considered as rejections given to an applicant with an *h* index higher than or equal to the median of the group of awarded applicants (under-estimation of an applicant's performance). Type I and type II errors can be determined with reference to both ex ante and ex post *h* index values. We use the word "error" here because of the statistical usage of this terminology (see Bornmann and Daniel, 2007a), but one should consider this value-neutral as deviations of the committee decisions from the scientometric prediction. Comparison of the ex ante and ex post meta-evaluations enables us to estimate how often the committee "picked the winners" (or not) by deviating from the scientometric prediction.

---

[1] In this example, 42 applications of psychologists were awarded in 2003, 46 in 2004, and 31 in 2005. The sample of corresponding best-rejected ones was equally stratified.

Bibliometric studies have demonstrated that factors other than scientific quality have a general influence on citation counts (Bornmann & Daniel, 2008): Citation counts are affected by the number of co-authors (Wuchty, Jones, & Uzzi, 2007) and the length of a paper (Bornmann & Daniel, 2007b) as well as the size of the citation window. This means that there is a positive association between citation counts and the number of co-authors and the size of a paper as well as the length of the citation window. By considering these factors in the statistical analysis, it becomes possible to establish a meaningful and adjusted co-variation between decisions made by the selection committee and the bibliometric data about the applicants.

We performed 16 multiple regression analyses (eight for each data set) with the statistical package Stata, which reveal the factors that exert a primary influence on citation counts. These models took the number of pages and the number of co-authors of each paper as covariates into account in addition to the decision variable (dichotomous variable: 0=rejected, 1=awarded). The publication years of the papers were included in the models predicting citation counts as exposure time (Long & Freese, 2006, pp. 370-372). We use the exposure option provided in Stata to take into account the time that a paper is available for citation. The violation of the assumption of independent observations by including citation counts of more than one paper per application was considered in the models by using the cluster option in Stata. This option specifies that the citation counts are independent across papers of different applicants, but are not necessarily independent within papers of the same applicant (Hosmer & Lemeshow, 2000, section 8.3).

The outcome variable (number of citations) in the models is a count variable. It indicates "how many times something has happened" (Long & Freese, 2006, p. 350). The Poisson distribution is often used to model information on counts. However, this distribution rarely fits in the statistical analysis of bibliometric data, due to overdispersion. "That is, the [Poisson] model underfits the amount of dispersion in the outcome" (Long & Freese, 2006, p. 372). Since the standard model to account for overdispersion is the negative binomial (Hausman, Hall, & Griliches, 1984), we calculated in the present study negative binomial regression models (Hilbe, 2007). According to Allison (1980) negative binomial fits scientific productivity distributions at best.

The regression analyses on citation counts in the present study are based on those applicants who published at least one paper (ex ante and ex post, respectively). Non-publishers had to be excluded from the analysis, because they had not published any paper that could have been cited. This might especially influence the MaWG results, as 8% (psychology) and 16% (economics) of the applicants had not published any papers ex ante and/ or ex post.

**Results**

*a. Awarded versus Rejected*

Table 2 shows the results of the Mann-Whitney test for the comparisons between awarded and (best-)rejected applicants of both agencies. If we consider the (119 + 251 =) 370 applications in psychology, the awarded applicants have on average an *h* index of 3.6 ex ante and 3.1 ex post, while the corresponding figures for the rejected applicants are 2.6 and 2.4, respectively. The difference between these two groups is statistically significant ex ante, but is not statistically significant ex post. If we cut off the tail of the distribution by only comparing the 119 awardees with the 119 best-rejected applications, the mean *h* index values of this latter group is 4.1 ex ante and 3.5 ex post. Both, ex ante and ex post the latter group has a higher performance than the awarded group of applicants. As the further figures in Table 2 for psychology show, this pattern of group differences holds true in case of the other performance indicators and holds also true for the different comparisons in economics. With regards to the statistical significance of the differences between the groups we found different results for ex ante and ex post comparisons. Most of the comparisons are statistically significant ex ante, but not ex post.

[Table 2 about here.]

Whereas the MaGW awardees have performance scores on average higher than the rejected applicants, the best-rejected subgroup has higher scores on average than the awardees. According to this latter result we found also a considerable amount of "erroneous" decisions. As Table 2 shows between 41 and 58% of the decisions in psychology and economics can be categorized as a type I or type II error if the ex ante performance is used as criterion. These percentages increase when the ex post performance is used (with one exception): Between 52 and 63% of the decisions in both disciplines show a type I or type II error. That means a considerable amount of applications could have been funded although they were rejected and vice versa – if the $h$ index is used as a validity criterion. This finding accords with that previously reported by Van den Besselaar and Leydesdorff (2007, 2009), although we use here different performance indicators. This raises the question of whether the results are specific for the social sciences or if they hold for the life sciences too.

The EMBO data set informs us that the MaGW results are partly transferable (see Table 2). Ex ante and ex post all parameters for the rejected applicants are lower than those for the awardees. Most of the differences between both groups are statistically significant. These results are in accordance with the results published in Bornmann, Wallon, and Ledin (2008a, 2008b) and point to the basic convergent and predictive validity of the EMBO selection decisions. With regards to the comparison of the awarded and the best-rejected applicants one has to differentiate between productivity and citation impact. In terms of productivity (mean number of publications and $h$ index) the awarded applicants perform better than the best-rejected applicants. (According to the results of Bornmann, Mutz, and Daniel (2008) the $h$ index is more a productivity than an impact indicator.) In terms of citation impact the situation is reversed. With one exception (YI programme, ex post) the best-rejected applicants perform better than the awarded applicants. The calculation of the error types shows that there are not only "erroneous" decisions among the rejected applicants (type II errors) as the comparison of the awarded and the best-rejected applicants indicate. Between 29 and 56% of the awarded applicants perform equal to or lower than an average rejected applicant.

In summary, the results of the Mann-Whitney test show that the MaGW and EMBO decisions are successful in removing the tails of the distributions in the applications, but less so in prioritizing among the top applicants. The advantage of the best-rejected applicants against awardees in the MaGW data, however, is not consistently found in the EMBO data. While the EMBO awardees outperform the best-rejected applicants in terms of productivity, with regard to citation impact the order is reversed. Negative binomial regression analysis will enable us to study the citation data in more detail and to find out whether the results hold true if factors are considered which have a general influence on citation counts.

*b. Negative binomial regression models*

Tables 3-6 summarize the results of the negative binomial regression models. The tables provide parameter estimates as exponents. For example, the number of co-authors of a publication in the case of all applicants in economics relates statistically significant to citations with a parameter estimate of 0.145 (see Table 3). This means that for each additional co-author, the odds of receiving citations increase by a factor of 1.16 (= exp(0.145)), holding all other variables in the model constant.

[Tables 3-6 about here.]

Similarly, there is a statistically significant association between funding decisions and citation counts in the case of all applicants in psychology with a parameter estimate of 0.222 ex ante, and this increases a bit to 0.275 ($p < 0.05$) ex post. These parameter values correspond to factors of 1.25 and 1.32, respectively, in the odds of receiving citations. The additional 25 and 32% of citation rates are provided in the bottom line of the respective tables 3 and 4. These calculations of the percent change in expected counts for a unit increase in the decision variable (from rejection to approval) following the regression models showed that being an awarded applicant increases the expected number of citations by 25% (ex ante) and 32% (ex post), respectively. As Tables 3 to 6 show in the comparison of the awarded and rejected applicants, the publications (ex ante and ex post) of the awarded applicants have statistically significant higher expected citation rates than those of the rejected applicants (published ex ante and ex post). These results confirm the basic convergent and predictive validity of both agencies, if citation impact is used as validity criterion. The one exception is economics for publications ex ante as well as ex post.

In the bottom line of the tables, one can read that the MaGW and EMBO selection decisions (0=rejected, 1=awarded) in all comparisons of the awardees with the best-rejected applicants are *negatively* related to citation counts, ex ante and ex post. The results for psychologists (ex post) are the single exception. That means when comparing similar sets of awardees with the best-rejected applications, the results show that the committees of both agencies select candidates that score on average weaker than the best rejected ones in terms of citation impact. However, not all regression coefficients for the decision variable in the tables are statistically significant (four statistically significant results out of eight regression models).

**Conclusions and discussion**

In this study we compared the selection decisions performed by two agencies for the selection of post doctoral fellows (EMBO), young investigators (MaGW and EMBO), and research grants (MaGW). The results of the statistical analyses show that the mean productivity and the mean citation impact of awarded applicants are higher prior and subsequent to application than the mean impact of rejected applicants. That means, there is a statistically significant association between selection decisions and the applicants' scientific achievements, if citation impact is used as a criterion for scientific achievement.

Measured against both criteria, the selection decisions are convergent and predictively valid.

In further analyses we tested whether this conclusion can be hold if certain subgroups of the applicants are compared. With the bibliometric data of the applicants prior and subsequent to application we compared the awarded and the best-rejected applicants. Additionally the extent of differences between the decision taken and an alternative decision based on past performance indicators was calculated in terms of type I and type II errors. The results are rather different: First, we found a high performance of the best-rejected applicants in all disciplines. Measured against the impact criterion, the selection decisions are *not convergent and predictive valid*. Measured against the applicants' productivity the results are heterogeneous and we found validity in the case of EMBO only, but not in the MaGW case. Second, we found that nearly one third (or more) of all applicants would not have been funded based on their performance only (type I error), and that nearly the same rate (or more) of applicants would have been funded as they had a high performance (type II error).

Our review of the literature revealed that other studies on peer review also report the occurrence of errors of this kind in selection decisions. Bornmann and Daniel (2007a) investigated the validity of decisions for awarding long-term fellowships to post doctoral researchers as practiced by the Boehringer Ingelheim Fonds (BIF). Approximately, one third of the decisions to award a fellowship to an applicant show a type I error, and about one third of the decisions not to award a fellowship to an applicant show a type II error.

Thorngate, Faregh, and Young (2002) comments as follows on the grants peer review of the Canadian Institutes of Health Research (CIHR, Ottawa): "Some of the losing proposals are truly bad, but not all; many of the rejected proposals are no worse than many of the funded ones … When proposals are abundant and money is scarce, the vast majority of putative funding errors are exclusory; a large number of proposals are rejected that are statistically indistinguishable from an equal number accepted" (p. 3).

All in all, our results are in part counter-intuitive. One would expect selection committees to be successful in "picking the winners." In terms of the applicants' (past and post) performance only (measured by bibliometric data), this seems not always to be the case. How can we understand these findings?

1. The most obvious one is in the nature of decision making about grant and fellowship applications. After removing the weaker applicants and applications, the performance of the applicants may play an important role in deciding on whom to fund, but at most as one of the criteria. Quality of the proposal plays a role, but also thematic criteria, and considerations of societal relevance. Above that, decision makers may want to increase the number of female researchers and researchers from minority groups, and therefore give those a preference within the group of good researchers. The scientific merit of the *proposed research* is expected to be a major criterion for awarding a grant. It is a limitation of our research that we did not take this into account, as the focus of this study is on the quality of the *applicants*. Applicants with lower calculated bibliometric measures may have proposed ideas with higher scientific merit and highly regarded authors may have proposed less innovative research.[2] There may or may not be a correlation between author quality and scientific merit of proposals, the strength or weakness of that correlation leaves room for the observed outcome.

2. If the different selection criteria used by the agencies correlate moderately at best[3], it is intuitively easy to understand that the best scoring non-successful group may have a higher average score than the successful group in *certain single dimensions* (e.g., citation impact). However, as the decision is based on a multi-criteria evaluation, the successful group may on average show a better composite score profile.

3. In general, papers published in journals covered by Thomson Reuters play an explicit and important role in the EMBO selection process (and in the area of life sciences as a whole), whereas this is not so clearly the case in the MaGW selection process (and in the area of social sciences as a whole). This may explain the differences of the results for the life sciences and the social sciences concerning the applicants' productivity.

Both of the agencies considered here are highly reputable and making serious efforts to organize the reviewing processes and careful selections thereupon. If only a few applications can be selected for funding (because of scarce resources), many applications of researchers with good past performance must be rejected. Using a more sociological and systemic perspective, we are inclined to think of the development of the sciences as self-organizing processes, and the question remains how funding systems effect the dynamics of science. In case of low approval rates combined with high rates of qualified applications, besides scientific criteria available resources and additional considerations decide on scientific advancement.

---
[2] In the MaGW case, the best rejected and awarded applicants did not differ in terms of average referee scores (Van den Besselaar and Leydesdorff 2007, 2009).
[3] This is also the case for e.g., the bibliometric indicators and the referee scores. In the MaGW case (Van den Besselaar and Leydesdorff, 2007) we found low correlations between past performance indicators and the referee scores. In the group of successful and best non-successful applicants this correlation was even zero: that means no *convergent validity*. Correlation between the referee score and the decision is low in both cases.

# References

Allison, P. D. (1980). Inequality and scientific productivity. *Social Studies of Science, 10*(2), 163-179.

Bornmann, L. & Daniel, H.-D. (2007a). Convergent validation of peer review decisions using the *h* index. Extent of and reasons for type I and type II errors. *Journal of Informetrics, 1*, 204–213.

Bornmann, L. & Daniel, H.-D. (2007b). Multiple publication on a single research study: does it pay? The influence of number of research articles on total citation counts in biomedicine. *Journal of the American Society for Information Science and Technology, 58*(8), 1100-1107.

Bornmann, L. & Daniel, H.-D. (2008). What do citation counts measure? A review of studies on citing behavior. *Journal of Documentation, 64*(1), 45-80.

Bornmann, L. & Daniel, H.-D. (2009). The state of *h* index research. Is the *h* index the ideal way to measure research performance. *EMBO Reports, 10*(1), 2-6.

Bornmann, L., Mutz, R., & Daniel, H.-D. (2008). Are there better indices for evaluation purposes than the *h* index? A comparison of nine different variants of the *h* index using data from biomedicine. *Journal of the American Society for Information Science and Technology, 59*(5), 830-837.

Bornmann, L., Wallon, G., & Ledin, A. (2008a). Does the committee peer review select the best applicants for funding? An investigation of the selection process for two European Molecular Biology Organization programmes. *PLoS ONE,* 3(10).

Bornmann, L., Wallon, G. & Ledin, A. (2008b). Is the *h* index related to (standard) bibliometric measures and to the assessments by peers? An investigation of the *h* index by using molecular life sciences data. *Research Evaluation, 17*(2), 149-156.

Daniel, H.-D., Mittag, S. & Bornmann, L. (2007). The potential and problems of peer evaluation in higher education and research. In: A. Cavalli (Ed.), *Quality assessment for higher education in Europe* (pp. 71-82). London, UK: Portland Press.

De Haan, J. (1994). *Research groups in Dutch sociology*. Amsterdam, the Netherlands: Thesis Publishers.

Elkana, Y., Lederberg, J., Merton, R. K., Thackray, A., & Zuckerman, H. (1978). *Toward a metric of science: the advent of science indicators*. New York, NY: Wiley.

Frey, B. S. & Osterloh, M. (2002). *Successful management by motivation: balancing intrinsic and extrinsic incentives*. Berlin, Germany: Springer.

Garfield, E. (1972). Citation analysis as a tool in journal evaluation. *Science* 178(4060), 471-479.

Garfield, E. (1979). *Citation indexing – its theory and application in science, technology, and humanities*. New York, NY: Wiley.

Hausman, J., Hall, B. H., & Griliches, Z. (1984). Econometric models for count data with an application to the patents R and D relationship. *Econometrica, 52*(4), 909-938.

Hilbe, J. M. (2007). *Negative binomial regression*. Cambridge, UK: Cambridge University Press.

Hornbostel, S., Böhmer, S., Klingsporn, B., Neufeld, J. & von Ins, M. (2009). Funding of young scientist and scientific excellence. *Scientometrics*, *79*(1), 171-190.

Hosmer, D. W. & Lemeshow, S. (2000). *Applied logistic regression* (2. ed.). Chichester, UK: Wiley.

Long, J. S. & Freese, J. (2006). *Regression models for categorical dependent variables using Stata* (2. ed.). College Station, TX: Stata Press, Stata Corporation.

Martin, B. R. & Irvine, J. (1983). Assessing basic research: some partial indicators of scientific progress in radio astronomy. *Research Policy,* 12, 61-90.

Martin, B. R. & Irvine, J. (1985). Evaluating the evaluators. *Social Studies of Science,* 15, 558-585.

Melin, G. & Danell, R. (2006). The top eight percent: development of approved and rejected applicants for a prestigious grant in Sweden. *Science and Public Policy*, 33(10), 702-712.

Merton, R. K. (1968). The Matthew effect in science. *Science,* 159, 56-63.

Merton, R. K. (1979). Foreword. In *E. Garfield, Citation indexing: its theory and application in science, technology and humanities* (pp. v-ix). New York, NY: Wiley.

Moed, H. F., Burger, W. J. M., Frankfort, J. G., & Van Raan, A. F. J. (1985). The use of bibliometric data for the measurement of university research performance. *Research Policy 14*, 131-149.

Mulkay, M. J. (1976). The mediating role of the scientific elite. *Social Studies of Science, 6*(34), 445-470.

Narin, F. (1976). *Evaluative bibliometrics: the use of publication and citation analysis in the evaluation of scientific activity*. Washington, DC: National Science Foundation.

Nederhof, A. J., & Van Raan, A. F. J. (1987). Peer review and bibliometric indicators of scientific performance: a comparison of cum laude doctorates with ordinary doctorates in physics. *Scientometrics, 11*(5), 333-350.

Straub, D. W. (2008a). Thirty years of service to the IS profession: time for renewal at MISQ? *MIS Quarterly*, *32*(1), iii-viii.

Straub, D. W. (2008b). Type II reviewing errors and the search for exciting papers. *MIS Quarterly*, *32*(2), V-X.

Thorngate, W., Faregh, N., & Young, M. (2002). Mining the archives: analyses of CIHR research grant applications. Retrieved April 28, 2005, from http://http-server.carleton.ca/~warrent/reports/mining_the_archives.pdf

Van den Besselaar, P. & Leydesdorff, L. (2007). *Past performance as predictor of successful grant applications*. Den Haag, the Netherlands: Rathenau Instituut.

Van den Besselaar, P., & Leydesdorff, L. (2009). Past performance, peer review, and project selection: a case study in the social and behavioral sciences. *Research Evaluation,* in print.

Wuchty, S., Jones, B. F., & Uzzi, B. (2007). The increasing dominance of teams in production of knowledge. *Science, 316*(5827), 1036-1039.

Zuckerman, H., & Merton, R. K. (1971). Patterns of evaluation in science: Institutionalisation, structure and functions of the referee system. *Minerva, 9*(1), 66-100.

**Acknowledgements**

The authors would like to thank Dr. Gerlind Wallon, deputy director of the European Molecular Biology Organization (EMBO), and Dr. Anna Ledin, working as a scientific secretary for the Royal Swedish Academy of Sciences in Stockholm (and former at EMBO), for providing the bibliographic data on the applicants to the EMBO Long Term Fellowship and Young Investigator Programmes. The authors also would like to thank the staff members of the Dutch Economics and Social Research Council (MaGW-NOW) for providing and preparing the data on the MaGW career grants and the MaGW open competition grants.23

Table 1. Number of applications and number of publications for the awarded, rejected and the best-rejected groups in the MaGW and EMBO data sets

|  | Total | Awarded | Rejected | Best-rejected |
|---|---:|---:|---:|---:|
| MaGW data |  |  |  |  |
| Psychology | 370 | 119 | 251 | 119 |
| Publications |  |  |  |  |
| ex ante | 2165 | 818 | 1347 | 895 |
| ex post | 3276 | 1246 | 2030 | 1347 |
| Economics | 301 | 46 | 255 | 46 |
| Publications |  |  |  |  |
| ex ante | 754 | 160 | 594 | 175 |
| ex post | 1020 | 173 | 847 | 209 |
| EMBO data |  |  |  |  |
| LTF | 668 | 130 | 538 | 130 |
| Publications |  |  |  |  |
| ex ante | 2227 | 586 | 1641 | 439 |
| ex post | 2320 | 539 | 1781 | 473 |
| YI | 297 | 39 | 258 | 39 |
| Publications |  |  |  |  |
| ex ante | 2153 | 313 | 1840 | 256 |
| ex post | 2292 | 403 | 1889 | 258 |

Table 2. Mean number of publications, mean number of total citations counts, mean *h* index values, and % type I and type II errors for the applications in the MaGW and EMBO data sets (ex ante and ex post)

*MaGW data set (*ex ante*)*

| Applications | | Mean number of publications | | Mean number of total citation counts | | Mean *h* index values | | % type I and type II errors | |
|---|---|---|---|---|---|---|---|---|---|
| Awarded | Rejected | Awarded | Rejected | Awarded | Rejected | Awarded | Rejected | Type I | Type II |
| Psychology | | | | | | | | | |
| 119 | 251 | 6.9* | 5.4* | 65.6* | 41.2* | 3.6* | 2.6* | 41 | 41 |
| 119 | 119 (best) | 6.9 | 7.5 | 65.6* | 76.5* | 3.6 | 4.1 | | |
| Economics | | | | | | | | | |
| 46 | 255 | 3.5* | 2.3* | 14.5* | 7.7* | 1.8* | 1.1* | 54 | 58 |
| 46 | 46 (best) | 3.5 | 3.8 | 14.5* | 25.5* | 1.8* | 2.5* | | |

*MaGW data set (*ex post*)*

| Applications | | Mean number of publications | | Mean number of total citation counts | | Mean *h* index values | | % type I and type II errors | |
|---|---|---|---|---|---|---|---|---|---|
| Awarded | Rejected | Awarded | Rejected | Awarded | Rejected | Awarded | Rejected | Type I | Type II |
| Psychology | | | | | | | | | |
| 119 | 251 | 6.8 | 5.9 | 65.3 | 41.6 | 3.1 | 2.4 | 53 | 52 |
| 119 | 119 (best) | 6.3 | 8.2 | 59.6 | 71.3 | 3.1 | 3.5 | | |
| Economics | | | | | | | | | |
| 46 | 255 | 3.3 | 2.8 | 12.9 | 11.0 | 1.4 | 1.2 | 63 | 56 |
| 46 | 46 (best) | 3.3 | 3.6 | 12.9 | 20.4 | 1.4 | 1.7 | | |

*EMBO data set (*ex ante*)*

| Applications | | Mean number of publications | | Mean number of total citation counts | | Mean $h$ index values | | % type I and type II errors | |
|---|---|---|---|---|---|---|---|---|---|
| LTF programme | | | | | | | | | |
| Awarded | Rejected | Awarded | Rejected | Awarded | Rejected | Awarded | Rejected | Type I | Type II |
| 130 | 538 | 4.5* | 3.1* | 267.5* | 111.5* | 4.1* | 2.7* | 29 | 26 |
| 130 | 130 (best) | 4.5* | 3.4* | 267.5* | 278.2* | 4.1* | 3.3* | | |
| YI programme | | | | | | | | | |
| 39 | 258 | 8.0 | 7.1 | 358.1* | 256.1* | 6.3 | 5.5 | 49 | 44 |
| 39 | 39 (best) | 8.0 | 6.6 | 358.1* | 632.7* | 6.3 | 6.0 | | |

*EMBO data set (*ex post*)*

| Applications | | Mean number of publications | | Mean number of total citation counts | | Mean $h$ index values | | % type I and type II errors | |
|---|---|---|---|---|---|---|---|---|---|
| LTF programme | | | | | | | | | |
| Awarded | Rejected | Awarded | Rejected | Awarded | Rejected | Awarded | Rejected | Type I | Type II |
| 130 | 538 | 4.2* | 3.3* | 171.1* | 111.9* | 3.6* | 2.9* | 56 | 52 |
| 130 | 130 (best) | 4.2 | 3.6 | 171.1 | 178.5 | 3.6 | 3.3 | | |
| YI programme | | | | | | | | | |
| 39 | 258 | 10.3* | 7.3* | 196.5* | 114.1* | 6.4* | 4.4* | 36 | 26 |
| 39 | 39 (best) | 10.3* | 6.6* | 196.5 | 161.2 | 6.4* | 4.7* | | |

* $p < 0.05$

**Table 3.** *MaGW data set (*ex ante*):* Negative binomial regression models predicting citations for papers published prior to application

|  | Psychology: All applicants | Psychology: Awarded and the best-rejected applicants | Economics: All applicants | Economics: Awarded and the best-rejected applicants |
|---|---|---|---|---|
| Decision (1=awarded) | 0.222* (2.45) | -0.0564 (-0.65) | 0.250 (1.46) | -0.463* (-2.85) |
| Number of Pages | 0.00825* (2.09) | 0.00916* (2.35) | 0.00167 (0.24) | -0.00682 (-0.91) |
| Number of co-authors | 0.0823* (3.66) | 0.0655* (3.16) | 0.145* (2.20) | 0.100 (1.44) |
| Publication year | (exposure) | (exposure) | (exposure) | (exposure) |
| Intercept | -6.003* (-49.94) | -5.674* (-49.19) | -6.840* (-25.64) | -5.856* (-19.35) |
| $n_{papers}$ | 2165 | 1713 | 754 | 335 |
| $n_{applicants\ (clusters)}$ | 319 | 225 | 206 | 82 |
| Papers per applicant (cluster) | minimum=1 mean=6.8 maximum=62 | minimum=1 mean=7.6 maximum=62 | minimum=1 mean=3.7 maximum=29 | minimum=1 mean=4.1 maximum=14 |
| Percent change in expected counts for a unit increase in "Decision" | 25% | -6% | 28% | -37% |

*Note.* ML-point estimates (the results of the *z*-test in parentheses).
* $p < 0.05$.
There are differences in number of papers between this table and Table 1, because some publications could not be included in the regression models because of missing values in number of pages and/ or number of co-authors. Furthermore, some applicants could not be included in the models, because they have no publications within the publication years considered here.

**Table 4.** *MaGW data set (*ex post*):* Negative binomial regression models predicting citations for papers published subsequent to application

|  | Psychology: All applicants | Psychology: Awarded and the best-rejected applicants | Economics: All applicants | Economics: Awarded and the best-rejected applicants |
|---|---|---|---|---|
| Decision (1=awarded) | 0.275* (2.76) | 0.0674 (0.69) | 0.0239 (0.13) | -0.336 (-1.48) |
| Number of Pages | 0.00985 (1.72) | 0.00896 (1.57) | 0.0201* (2.70) | 0.0252* (2.56) |
| Number of co-authors | 0.0294* (2.06) | 0.0207 (1.52) | 0.0314 (1.33) | -0.0560 (-1.08) |
| Publication year | (exposure) | (exposure) | (exposure) | (exposure) |
| Intercept | -6.000* (-63.31) | -5.746* (-58.76) | -6.772* (-34.36) | -6.285* (-16.08) |
| $n_{papers}$ | 3276 | 2593 | 1020 | 382 |
| $n_{applicants\ (clusters)}$ | 322 | 219 | 235 | 82 |
| Papers per applicant (cluster) | minimum=1 mean=10.2 maximum=93 | minimum=1 mean=11.8 maximum=93 | minimum=1 mean=4.3 maximum=29 | minimum=1 mean=4.7 maximum=17 |
| Percent change in expected counts for a unit increase in "Decision" | 32% | 7% | 2% | -29% |

*Note.* ML-point estimates (the results of the *z*-test in parentheses).
* $p < 0.05$.
There are differences in number of papers between this table and Table 1, because some publications could not be included in the regression models because of missing values in number of pages and/ or number of co-authors. Furthermore, some applicants could not be included in the models, because they have no publications within the publication years considered here.

**Table 5**. *EMBO data set (ex ante)*: Negative binomial regression models predicting citations for papers published prior to application

|  | LTF programme: All applicants | LTF programme: Awarded and the best-rejected applicants | YI programme: All applicants | YI programme: Awarded and the best-rejected applicants |
|---|---|---|---|---|
| Decision (1=awarded) | 0.467* (5.24) | -0.347* (-3.81) | 0.281* (2.49) | -0.678* (-5.69) |
| Number of Pages | 0.0366* (5.16) | 0.0194* (2.57) | 0.0401* (3.45) | 0.0232 (1.57) |
| Number of co-authors | 0.0321 (1.61) | 0.0176 (1.91) | 0.0596* (4.57) | 0.0574* (3.08) |
| Publication year | (exposure) | (exposure) | (exposure) | (exposure) |
| Intercept | -4.504* (-31.72) | -3.460* (-32.56) | -4.746* (-36.30) | -3.644* (-19.73) |
| $n_{papers}$ | 2221 | 1022 | 2145 | 567 |
| $n_{applicants\ (clusters)}$ | 634 | 257 | 291 | 78 |
| Papers per applicant (cluster) | minimum=1 mean=3.5 maximum=17 | minimum=1 mean=4 maximum=15 | minimum=1 mean=7.4 maximum=42 | minimum=1 mean=7.3 maximum=29 |
| Percent change in expected counts for a unit increase in "Decision" | 60% | -29% | 32% | -49% |

*Note*. ML-point estimates (the results of the *z*-test in parentheses).
* $p < 0.05$.
There are differences in number of papers between this table and Table 1, because some publications could not be included in the regression models because of missing values in number of pages and/ or number of co-authors. Furthermore, some applicants could not be included in the models, because they have no publications within the publication years considered here.

**Table 6**. *EMBO data set (ex post)*: Negative binomial regression models predicting citations for papers published subsequent to application

|  | LTF programme: All applicants | LTF programme: Awarded and the best-rejected applicants | YI programme: All applicants | YI programme: Awarded and the best-rejected applicants |
|---|---|---|---|---|
| Decision (1=awarded) | 0.187* (2.02) | -0.181 (-1.70) | 0.295* (2.67) | -0.158 (-1.13) |
| Number of Pages | 0.0164* (2.63) | 0.00910 (1.02) | 0.0173 (1.76) | 0.0147 (1.28) |
| Number of co-authors | 0.0143 (0.80) | 0.00465 (1.03) | 0.0320 (1.76) | 0.0379* (2.31) |
| Publication year | (exposure) | (exposure) | (exposure) | (exposure) |
| Intercept | -4.323* (-31.83) | -3.826* (-32.90) | -5.262* (-29.94) | -4.816* (-25.86) |
| $n_{papers}$ | 2306 | 1004 | 2236 | 642 |
| $n_{applicants\ (clusters)}$ | 607 | 242 | 288 | 77 |
| Papers per applicant (cluster) | minimum=1 mean=3.8 maximum=18 | minimum=1 mean=4.1 maximum=18 | minimum=1 mean=7.8 maximum=40 | minimum=1 mean=8.3 maximum=40 |
| Percent change in expected counts for a unit increase in "Decision" | 21% | -17% | 34% | -15% |

*Note.* ML-point estimates (the results of the *z*-test in parentheses).
* $p < 0.05$.
There are differences in number of papers between this table and Table 1, because some publications could not be included in the regression models because of missing values in number of pages and/ or number of co-authors. Furthermore, some applicants could not be included in the models, because they have no publications within the publication years considered here.